\makeatletter
\disable@package@load{breakurl}{}
\makeatother
\documentclass[sn-mathphys-num]{sn-jnl}

\usepackage[english]{babel}
\usepackage{graphicx}
\makeatletter
\def\maxwidth{\ifdim\Gin@nat@width>\linewidth\linewidth\else\Gin@nat@width\fi}
\def\maxheight{\ifdim\Gin@nat@height>\textheight\textheight\else\Gin@nat@height\fi}
\makeatother
\setkeys{Gin}{width=\maxwidth,height=\maxheight,keepaspectratio}

\makeatletter
\def\fps@figure{htbp}
\makeatother

\usepackage{amsmath}
\usepackage{amssymb}
\usepackage{mathtools}
\usepackage{physics}
\usepackage{color,soul}
\usepackage{siunitx}
\usepackage{multirow}
\DeclareSIUnit\gauss{G}

\usepackage{upgreek}
\usepackage[dvipsnames]{xcolor}
\usepackage{microtype}
\usepackage{xcolor}

\graphicspath{{Figs/}}
\begin{document}
\title{Production and magnetic self-confinement of $e^-e^+$ plasma by an extremely intense laser pulse incident on a structured solid target}

\date{\today}

\author[1]{\fnm{Alexander} \sur{Samsonov}}
\author*[1]{\fnm{Alexander} \sur{Pukhov}}
\email{pukhov@tp1.hhu.de}
\affil[1]{, \orgname{Institute for Theoretical Physics I, Heinrich Heine University Düsseldorf, 40225 Düsseldorf, Germany}}

\abstract
    {We propose an all-optical, single-laser-pulse scheme for generating dense, relativistic, strongly-magnetized electron-positron pair plasma. The scheme involves the interaction of an extremely intense ($I \gtrsim \SI{e24}{\watt/\cm^2}$) circularly polarized laser pulse with a solid-density target containing a conical cavity. Through full-scale three-dimensional particle-in-cell (PIC) simulations that account for quantum electrodynamical effects, it is shown that this interaction results in two significant outcomes: first, the generation of quasi-static axial magnetic fields reaching tens of gigagauss due to the inverse Faraday effect; and second, the production of large quantities of electron-positron pairs (up to $\num{e13}$) via the Breit-Wheeler process. The $e^-e^+$ plasma becomes trapped in the magnetic field and remains confined for hundreds of femtoseconds, far exceeding the laser timescale. The dependency of pair plasma parameters, as well as the efficiency of plasma production and confinement, is discussed in relation to the properties of the laser pulse and the target. Realizing this scheme experimentally would enable the investigation of physical processes relevant to extreme astrophysical environments.}
\maketitle


Electron-positron ($e^-e^+$) plasmas are found in some of the most extreme environments in our Universe, such as near black holes, neutron stars, and quasars.
These astrophysical objects are very distant and cannot be directly probed, leading to a limited amount of empirical data about the physical processes occurring in their vicinity.
To better understand and test the physics of these processes, recreating similar conditions in a controlled laboratory setting is crucial.

Significant efforts are currently underway to develop experimental facilities capable of generating such extreme conditions. A common method for producing electron-positron pairs involves directing a high-energy electron beam from a conventional accelerator through a thick high-Z material. The electrons can either generate high-energy photons via bremsstrahlung~\cite{tsaiPairProductionBremsstrahlung1974}, which subsequently create electron-positron pairs through the Bethe-Heitler process~\cite{bethe1934stopping}, or produce pairs directly through the trident process~\cite{Baier98}. This method is advantageous due to its simplicity and minimal additional setup requirements.

Although this straightforward configuration generates a large number of positrons, it is insufficient for the formation of a true pair plasma. For a plasma to exhibit collective behavior, the size of the electron-positron pair cloud must exceed the Debye radius. Additionally, for the pair plasma to be long-lived, the particles need to be confined, such as by a strong magnetic field. Previous attempts using this approach often resulted in low electron-positron beam density, which prevented the observation of plasma-like behavior. Only recently has it been demonstrated that sufficient pairs can be generated in these interactions to observe collective plasma effects, although long-term confinement has not yet been achieved~\citep{arrowsmithLaboratoryRealizationRelativistic2024}.

A promising alternative to conventional accelerators is offered by multi-PW laser systems, such as ELI \citep{ELI}, Apollon \cite{Apollon}, XCELS \citep{XCELS-HPLSE}, SEL \citep{SEL, SEL2}, and SULF \citep{SULF}.
These lasers generate ultrashort tens of femtoseconds level EM radiation with an optical wavelength $\lambda\sim\SI{1}{\um}$, which when focused can reach extreme intensities exceeding $\SI{e24}{\watt/\cm^2}$.
Field strength in the focus of such a laser pulse is commonly expressed in terms of its dimensionless amplitude $a_0$
\begin{equation}
  a_0 = \frac{e}{mc}\sqrt{-A_\mu A^\mu} \equiv \frac{eE_0}{m c \omega}\approx 0.85 \sqrt{I[\SI{e18}{\watt /\centi\meter^2}]}\lambda[\si{\micro\meter}],
\end{equation}
where $m$ and $e>0$ are the mass and the absolute value of the electron charge, respectively, $c$ is the speed of light, $A_\mu$ is the vector potential of the EM field, $ E_0$ is the field strength and $\omega$ is the central frequency.
The progress of the laser technology in the 20th century made it possible to implement Veksler's idea of coherent acceleration of particles~\cite{veksler1957principle} by generating high accelerating gradients in plasma during the propagation of intense laser radiation through it. Using short powerful laser pulses one can accelerate electrons to high energies over much shorter distances compared to the conventional accelerators~\cite{tajima1979laser, pukhovLaserWakeField2002, faure2004laser, esarey2009physics, clayton2010self, Kostyukov2015UFN, wenPolarizedLaserwakefieldacceleratedKiloampere2019, palastro2020dephasingless, tajima2020wakefield}.
In this context, the previously mentioned scheme for producing electron-positron pairs using a high-Z converter has been actively explored, with the primary modification being that the seed electron beam is accelerated through a laser-plasma interaction~\citep{sarriGenerationNeutralHighdensity2015, sarriOverviewLaserdrivenGeneration2015}.

\begin{figure}
\centering{ \includegraphics[width=0.5\linewidth]{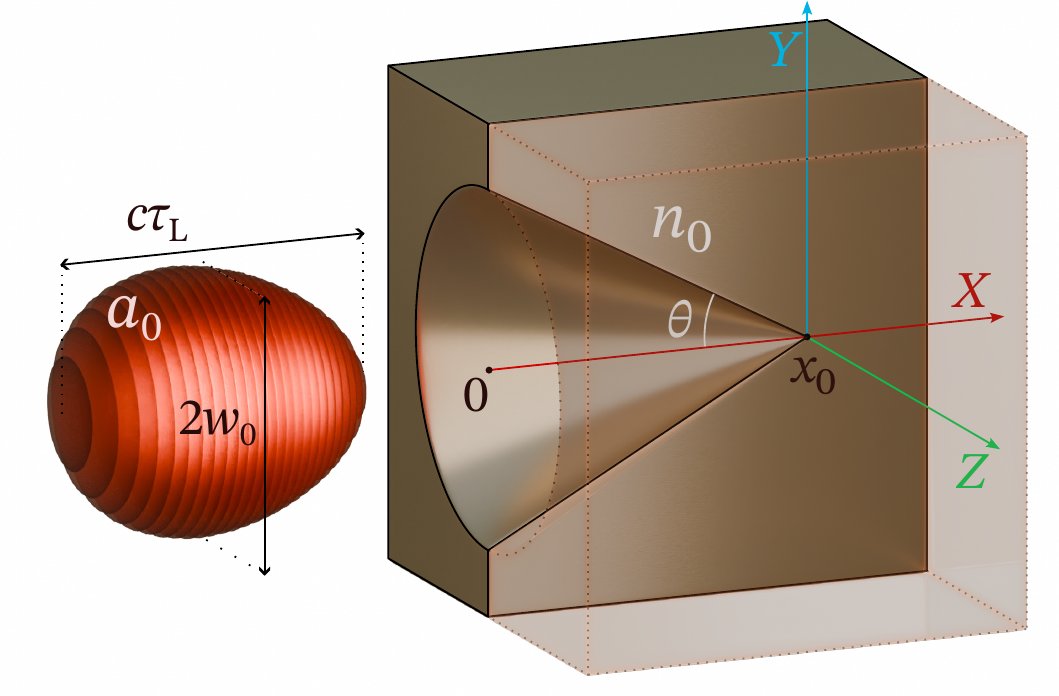} }
\caption{\label{fig-Setup} Simulation setup. The laser pulse is incident on a uniform target with a conical cavity.}
\end{figure}%

A different approach to producing pair plasma using relativistically strong laser pulses exploits nonlinear quantum electrodynamical (QED) effects during the interaction of such lasers and high-energy particles (with Lorentz factors 
$\gamma\gg 1$). Nonlinearity of the QED processes is governed by a Lorentz-invariant parameter $\chi$
\begin{equation}
  \chi = \frac{e\hbar\sqrt{-\left( F_{\mu\nu}p^\nu \right)^2}}{m^3c^4} = \frac{1}{E_\mathrm{S} mc} \sqrt{{\left( \frac{\varepsilon}{c}\vb{E} + \vb{p}\times\vb{B} \right)}^2 - {\left(\vb{p}\vb{E} \right)}^2} ,
\end{equation}
where $F_{\mu\nu}=\partial_\mu A_\nu - \partial_\nu A_\mu$is the EM field tensor, $\varepsilon$ and $\vb{p}$ are energy and momentum of the particle respectively, $E_\mathrm{S} = m^2 c^3/e\hbar\approx\SI{1.32e18}{\volt/\m}$ is a critical field of QED or Sauter-Schwinger field~\cite{Berestetskii82, Baier98}, $\hbar$ is Planck's constant.
This expression can be written for photons in an identical way, taking into account that $\varepsilon=\hbar\omega$, $p^\nu=\hbar k^\nu$.
With the increase of $\chi$ the maximum of the spectrum of the synchrotron radiation of the electrons shifts towards the larger energies, i.e. the electrons can radiate individual hard photons, which carry away a large portion of the electron energy.
If value of $\chi$ for such a photon exceeds unity, it can decay into an electron-positron pair in the nonlinear Breit-Wheeler process.
Successive radiation of hard photons by the electrons and positrons and decay of the former into new pairs can lead to an avalanche-like growth of the total number of particles.
This process is called a QED or an electromagnetic cascade~\cite{elkinaQEDCascadesInduced2011}.
Extensive research has been carried out focused on using extremely intense lasers to generate dense clouds of electron-positron pairs  with notable progress reported in recent studies~\cite{nerushRadiationEmissionExtreme2007,bellPossibilityProlificPair2008,nerushLaserFieldAbsorption2011,ridgersDenseElectronPositronPlasmas2012,kirkPairPlasmaCushions2013,narozhnyQuantumelectrodynamicCascadesIntense2015,zhuDenseGeVElectron2016,kostyukovProductionDynamicsPositrons2016, grismayerSeededQEDCascades2017,jirkaQEDCascade102017,luoQEDCascadeSaturation2018,yuanSpatiotemporalDistributionsPair2018,sorboEfficientIonAcceleration2018,luEnhancedCopiousElectron2018,samsonov2019laser,efimenkoLaserdrivenPlasmaPinching2019,chenPerspectivesRelativisticElectron2023}. 

In this paper, we demonstrate that when the laser power exceeds a certain threshold, the number of generated electron-positron pairs becomes so large that they begin to exhibit collective plasma behavior. Furthermore, the strong self-generated quasi-static magnetic fields, produced during the laser-solid target interaction, can trap and confine this plasma for hundreds of femtoseconds, allowing for the study of its collective properties.

In particular, we examine the interaction of a single circularly polarized laser pulse with a micro-structured, solid-density target. The target is a homogeneous slab with a cone-shaped volume removed along the laser's propagation axis.
The
laser pulse
effectively accelerates the electrons along the surface of the cone cavity and then reflects from the tip of the cavity.
The accelerated electrons collide with the reflected pulse leading to copious production of electron-positron pairs.
Additionally, the longitudinal flow of relativistic electrons generates a strong quasi-static azimuthal magnetic field.
As the relativistic pairs traverse this field, they rapidly lose energy due to the radiation reaction, becoming trapped inside the cone and confined for an extended period. The electron-positron plasma is then governed by magneto-hydrodynamics.

Interestingly, the electron-positron plasma remains effectively separated from the surrounding electron-ion plasma of the initial cone for hundreds of femtoseconds. This key feature distinguishes the proposed configuration from other experimental setups for studying strongly magnetized electron-positron plasma.

To investigate the production of electron-positron pair plasma in the interaction between an extremely intense laser pulse and a solid target with a conical cavity, a series of full-scale three-dimensional QED-PIC simulations were conducted. The general setup of these simulations is schematically depicted in Figure~\ref{fig-Setup} (with a 3D visualization of the reference simulation available in the supplemental material).


\begin{figure*}
\centering{\includegraphics{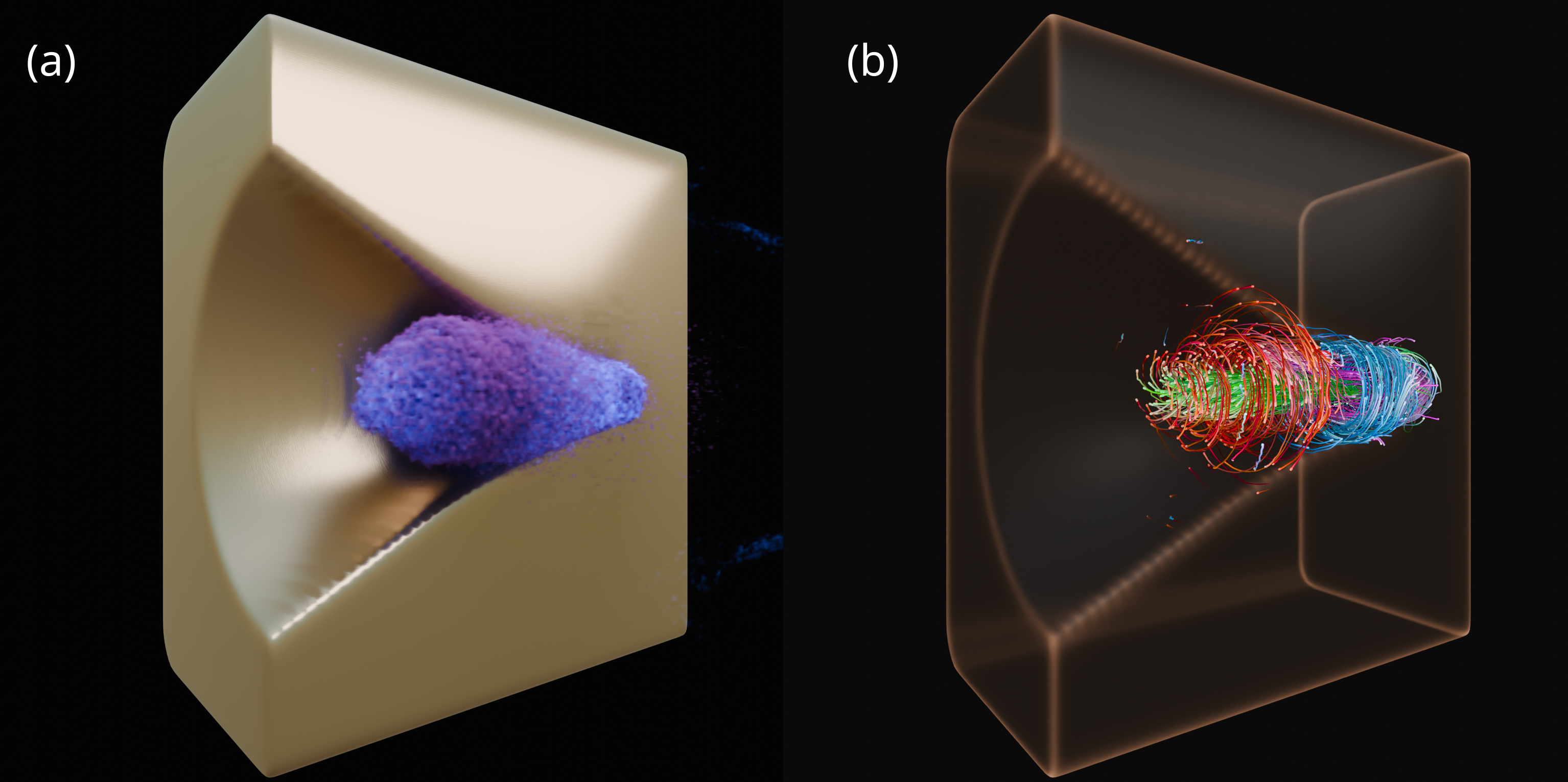}}
\caption{
  \label{fig-3d}
  Visualisation of (a) the positrons density (purple-blue cloud) and the surface of the constant ion density ($n=800n_\mathrm{cr}$) and (b) the pathlines of the quasi-static magnetic field in the reference simulation at $t = t_\mathrm{foc} + \SI{100}{\femto\s}$. The red (blue) color depicts predominance of the positive (negative) azimuthal component $B_\theta$, while the green (magenta) color depicts predominance of the positive (negative) axial component $B_x$. Density of the field lines corresponds to the magnitude of the magnetic field amplitude $|\vb{B}|$.
  }
\end{figure*}

\begin{figure*}
\centering{\includegraphics{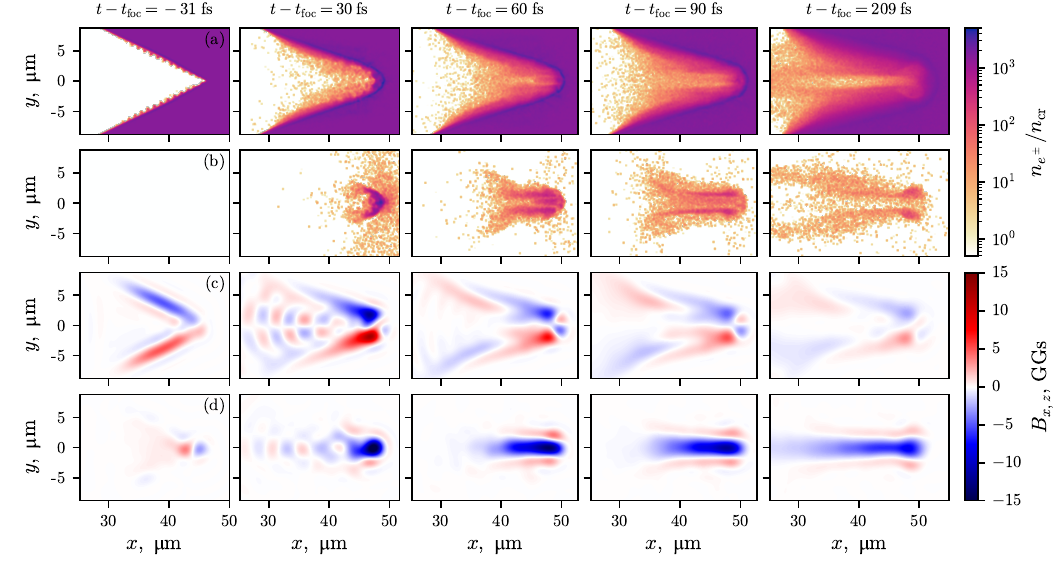}}
\caption{
  \label{fig-Density}
  Evolution of the (a) electron and (b) positron density and (c) $x$ and (d) $z$ components of the quasi-static magnetic field (see Methods) during the interaction in the reference simulation.
  The initial parameters of the simulation are reported in the text.
  }
\end{figure*}

In what follows we qualitatively describe the processes which are observed in a particular reference simulation with the target parameters ${n_0 = 1500 n_\mathrm{cr} \approx \SI{2e24}{\cm^{-3}}}$, $\tan\theta = 0.5$ (see Methods) and the laser parameters
$a_0 = 1000$, $w_0 = 3.5\lambda$, $\tau_\mathrm{L}= \SI{45.5}{\femto\s}$.
This corresponds to the peak intensity $\SI{1.65e24}{\watt /\centi\meter^2}$, total pulse energy $\SI{15}{\kilo\joule}$, and power $\SI{330}{\peta\watt}$.
The evolution of the fields and the particle densities in that simulation is shown in Figure~\ref{fig-Density}.
We roughly distinguish three main stages of the interaction, which follow one after another.
At the initial stage of the interaction (first column in Fig.~\ref{fig-Density}) the laser pulse extracts the electrons from the walls of the conical cavity and accelerates them along the surface, similarly to the case of a laser pulse interacting with a plane target at grazing angle \citep{serebryakovNearsurfaceElectronAcceleration2017,shen2024high}.
Previous research has also shown that abundant $e^-e^+$ pair production can be triggered by two counter-propagating laser pulses that are grazingly incident on a solid density target~\cite{filipovicQEDEffectsGrazing2022, samsonovGenerationElectronPositron2023}.

The axial electron current generates a strong azimuthal magnetic field, with its maximum intensity near the surface of the cavity. Additionally, since the laser is circularly polarized, the electrons undergo circular motion in the transverse plane. Due to the sharp boundary in electron density, a strong azimuthal current forms at the cavity's surface, which in turn generates an axial quasi-static magnetic field. This process is known as the inverse Faraday effect  \citep{pitaevskiiElectricForcesTransparent1961, shengInverseFaradayEffect1996, hainesGenerationAxialMagnetic2001, shvetsMagneticFieldGeneration2002, kostyukovMagneticfieldGenerationElectron2002,nuterGainElectronOrbital2020,jiangMagneticFieldAmplification2021,longmanKiloTeslaAxialMagnetic2021}.
As a result, during the initial stage of the interaction, powerful currents and corresponding quasi-static magnetic fields are produced, as illustrated in Fig.~\ref{fig-3d}~(b) and Video 2 in supplemental material.

At these high intensities, radiation reaction further contributes to the inverse Faraday effect \citep{liseykinaInverseFaradayEffect2016, liseykinaQuantumEffectsRadiation2021, liseykina2023inverse}.
This implies that even the interaction of an extremely intense laser with an initially homogeneous target (without a cavity) can produce a strong axial magnetic field, as demonstrated in prior studies \citep{samsonovEffectElectronPositron2021}.


In the next stage, the laser peak reaches the tip of the cavity at $t_\mathrm{foc} \equiv (x_\mathrm{L} - x_0) / c$, where $x_\mathrm{L}$ is the initial location of the center of the laser pulse.
The laser reflects backward, forming a standing wave for a brief period (second and third columns in Fig.~\ref{fig-Density}).
In this standing wave configuration, electrons experience stronger transverse forces compared to when they co-propagate with the laser pulse, leading to efficient emission of hard photons through nonlinear Compton scattering. These hard photons can subsequently decay into electron-positron pairs via the nonlinear Breit-Wheeler process. The laser pulse's duration and intensity are crucial in determining the total yield of pairs during this stage. In the simulation, the total number of pairs reaches $\num{1.8e13}$, while the density peaks at $\SI{5.5e23}{\cm^{-3}}$.
This high pair plasma density effectively absorbs a significant portion of the laser pulse.


The final stage of the interaction is marked by the laser pulse either exiting the interaction region or being fully absorbed by the pair plasma (as shown in the fourth and fifth columns of Fig. \ref{fig-Density}). During this phase, the energy distribution of the generated pairs transitions, and in the absence of laser fields, the energetic particles cool down via synchrotron radiation in the quasi-static magnetic fields. The pair energy distribution then rapidly relaxes into a Maxwell–Jüttner–Synge distribution~\cite{degrootRelativisticKineticTheory1980}, with a temperature of approximately 3 MeV in the simulation.

Despite the local anisotropies induced by the strong magnetic field, the overall angular distribution of the electron-positron pairs remains isotropic. At this stage, the plasma's evolution is entirely governed by magneto-hydrodynamics (MHD). The stability of similar plasma structures, particularly concerning various MHD instabilities, has been examined in previous studies, such as Refs. \cite{helanderGyrokineticStabilityTheory2016, istominStabilityRelativisticRotating1994, istominStabilityRelativisticRotating1996}.


The primary condition for the manifestation of collective plasma effects is that the plasma size,$L_\mathrm{pl}$ must significantly exceed both the Debye length, $\lambda_\mathrm{D}$and the skin depth, $l_\mathrm{s}$, which are defined as follows:
\begin{gather}
  l_\mathrm{s} \equiv \frac{c}{\omega_\mathrm{pl}} = \sqrt{\frac{ \langle \gamma_\pm \rangle m_e c^2}{8 \pi e^2 n_{e^\pm}}}, \\
  \lambda_\mathrm{D} \equiv \sqrt{\frac{ k_\mathrm{B} T_\pm }{8 \pi e^2 n_{e^\pm}}},
\end{gather}
where $n_{e^\pm}$ is the number density of pairs, hence the factor 2 in the plasma frequency definition.
Since the produced pairs remain ultrarelativistic for the whole duration of the simulation the ratio between the average pairs energy $\langle\gamma_\pm\rangle mc^2$ and their temperature $k_\mathrm{B} T_\pm$ equals to $3$, thus $l_\mathrm{s} = \sqrt{3} \lambda_\mathrm{D}$~\cite{wei-keDependenceAverageLorentz2005, stensonDebyeLengthPlasma2017}.
At the end of the simulation, corresponding to $t\approx t_\mathrm{foc} + \SI{200}{\femto\second}$ the plasma density settles at around $\SI{e23}{\cm^{-3}}$, which corresponds to the skin depth $l_\mathrm{s}\approx \SI{50}{\nano\m}$.
With the plasma scale estimated as $\SI{20}{\um}\times\SI{10}{\um}\times\SI{10}{\um}$ the condition $l_\mathrm{s}\ll L_\mathrm{pl}$ is satisfied with a large margin.

To characterize the magnetization of the plasma we estimate the Larmor radius
\begin{equation}
  r_\mathrm{L} = \frac{mc^2\langle \gamma_\pm \rangle}{eB}.
\end{equation}
At the end of the simulation, the magnetic field magnitude stabilizes at $\SI{6}{\giga\gauss}$, wesulting in a Larmor radius of $\SI{50}{\nm}$. This radius is significantly smaller than the plasma scale $L_\mathrm{pl}$, indicating that the produced plasma is strongly magnetized.



\begin{figure*}
  \centering{\includegraphics{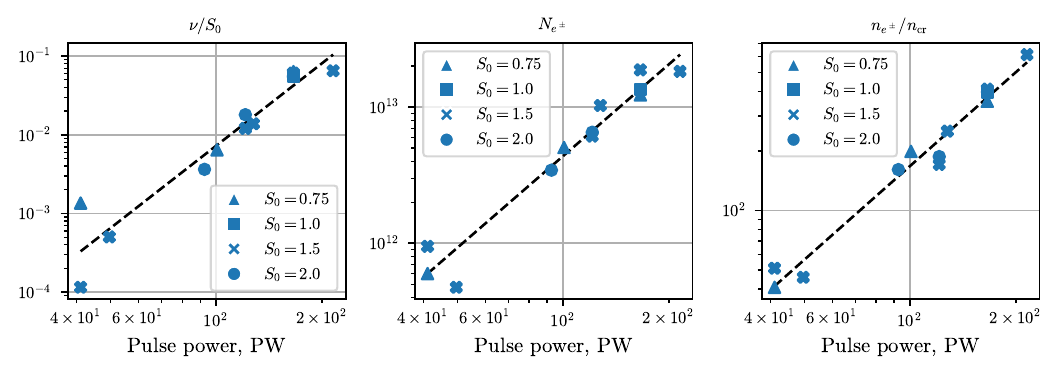}}
  \caption{\label{fig-Results} Dependency of the positrons trapping efficiency $\nu$, normalized to $S_0$, the total number of the produced $e^-e^+$ pairs $N_{e^\pm}$ and their maximum density $n_{e^\pm}$ on the pulse power.
  The black dashed lines indicate the linear fit in double logarithmic scale.}
\end{figure*}

We conducted a series of simulations, varying the parameters of the laser — specifically its strength $a_0$, duration $\tau_\mathrm{L}$ and beam waist $w_0$ -  as well as the target parameters, including its density $n_0$ and cone angle $\theta$.
The results are summarized in Fig.~\ref{fig-Results} and Table 1 in supplemental material.

We evaluated the resulting pair plasma parameters using several metrics: the total number of positrons $N_{e^\pm}$ and , the maximum positron density $n_{e^\pm}$ of the produced positrons, 
the fraction of trapped positrons $\nu$, and positron fraction in the plasma composition $\eta_{e^+}$, i.e. the ratio of the total number of positrons to the total number of all the particles: electrons, positrons and ions.
, which represents the ratio of the total number of positrons to the total number of all particles (electrons, positrons, and ions). The latter three metrics were calculated by averaging over the volume occupied by positrons (see Methods).

In the subset of simulations where the cone angle was constant $\tan\theta = 0.5$, the maximum correlation of these metrics was found with the laser pulse power $P$.
Specifically, $\nu/S_0$ scales as $P^3 $, where $S_0\equiv  n_0 / a_0 n_\mathrm{cr}$ is the relativistic similarity parameter~\cite{gordienko2005scalings}, $n_{e^\pm}$ scales as $P$, and $N_{e^\pm}$ scales as $ P^{2.2}$.

\begin{figure}
  \centering{\includegraphics{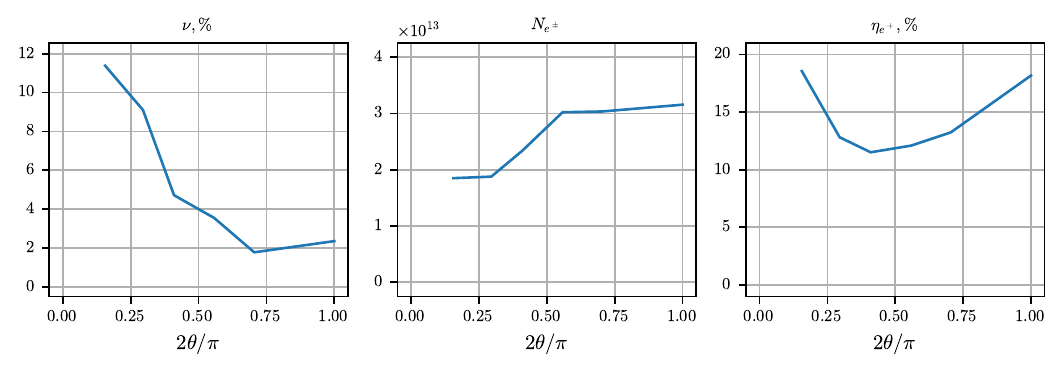}}
  \caption{\label{fig-Angle} Dependency of the trapping efficiency $\nu$, total number of produced $e^-e^+$ pairs $N_{e^\pm}$ and the positron fraction in the plasma composition $\eta_{e^+}$ on the cone angle $\theta$ in the simulations with the fixed target density $n_0 = 1500 n_\mathrm{cr} \approx \SI{2e23}{\cm^{-3}}$ and the laser parameters: $a_0 = 1000$, $w_0 = 3.5\lambda$, $\tau_\mathrm{L} = \SI{45.5}{\femto\second}$.}
\end{figure}

We have also performed a series of simulations with fixed laser parameters and density of the target, i.e. only changing the cone angle $\theta$.
Dependencies of the aforementioned metrics on $\theta$ are demonstrated in Fig.~\ref{fig-Angle}.
The Figure clearly demonstrates the advantage of using the structured target compared to the flat target ($\theta=\pi/2$) for reaching larger fraction of the positrons that are trapped in the self-generated magnetic field, albeit at the expense of the slightly decreased total positron yield.


The minimal requirements for the laser pulse at which significant trapping was still observed in the simulations are: energy $\SI{3}{\kilo\joule}$ at duration $\SI{16.2}{\femto\second}$ and $F/5.5$ focusing ($w_0 = 3.5\lambda$), corresponding to $a_0 = 750$, intensity $I=\SI{9.3e23}{\watt/\cm^2}$ and power $P=\SI{186}{\peta\watt}$.
Such parameters should be potentially achievable at the soon to come laser facilities, such as XCELS~\citep{XCELS-HPLSE}, SEL~\citep{SEL} and SULF~\citep{SULF}.


In conclusion, we have demonstrated an all-optical single-laser beam setup capable of producing dense, relativistic, strongly magnetized electron-positron plasma with a spatial scale significantly larger than its skin depth and Larmor radius, and a lifetime exceeding several hundred femtoseconds. The simulations indicate the potential to produce plasma with a total positron yield exceeding $\num{e13}$, with more than $10\%$ confined by the self-generated magnetic field,  a peak density surpassing 
 $\SI{e24}{\cm^{-3}}$and a laser energy conversion efficiency of over $6\%$ using laser parameters expected from future sub-exawatt laser facilities.
 
This scheme addresses three critical challenges in producing laboratory pair plasma. First, the production of high-energy seed particles is achieved by interacting the laser pulse with a target at a grazing angle, resulting in efficient electron acceleration. Second, the conversion of seed particles into dense pair plasma is facilitated by the formation of a standing wave structure, where the accelerated electrons trigger a QED cascade. Finally, the confinement of the produced plasma for a sufficiently long duration—a less explored challenge—is resolved by the generation of strong quasi-static magnetic fields through the inverse Faraday effect. This is a significant advantage, as plasmas in astrophysical environments are typically strongly magnetized.
Thus, the proposed scheme offers a versatile tool for investigating the properties of pair plasma across a wide range of parameters, making it particularly valuable for simulating extreme astrophysical environments in a laboratory setting.

The work is supported by BMBF-Project: 05P24PF1 and DFG project PU 213/9-1.
The authors gratefully acknowledge the Gauss Centre for Supercomputing e.V.~\cite{Juwels} for funding this project (lpqed) by providing computing time through the John von Neumann Institute for Computing (NIC) on the GCS Supercomputer JUWELS at Jülich Supercomputing Centre (JSC). 

\section*{Methods}
\label{methods}

\subsection*{3D QED-PIC simulations}
The simulations were performed using the VLPL code \citep{pukhovParticleInCellCodesPlasmabased2016}.
A novel implementation of the Monte-Carlo QED module is based on the inversion sampling technique similar to the method proposed in \citet{volokitinOptimizedEventGenerator2023}.
This method allows to decouple the regular PIC loop with a constant time step required to resolve plasma and/or laser processes from the QED loop which, depending on the dynamic parameter $\chi$, may require significantly smaller time steps.
Moreover, spin-resolved probabilities \citep{songDensePolarizedPositrons2022, zhuangLaserdrivenLeptonPolarization2023} are implemented in the algorithm.
A more detailed description of the algorithm will be published separately.

In simulation the target is initialized as a fully ionized plasma with initially 8 electrons and 4 ions per numerical cell.
The ions mass to charge ratio is equal to double that for proton.
The initial density distribution of electrons is governed by the expression
\begin{equation}
  \nonumber
  n = \begin{cases}
    n_0 & \text{if } x > 0 \land \rho > (x_0 - x) \tan\theta,  \\
    0 & \text{otherwise},
  \end{cases}
\end{equation}
where \(\rho = \sqrt{y^2 + z^2}\) is the distance from the symmetry axis.
The value of $n_0$ in the performed simulations was on the order of $\SIrange{e23}{e24}{\cm^{-3}}$ which generally corresponds to the density of solids.
The laser radiation is generated at the left boundary of the simulation box to produce a gaussian pulse focused to the tip of the conical cavity with the waist $w_0$, the amplitude at the focus $a_0$ and the duration $\tau_\mathrm{L}$.
In all simulations the depth and the diameter of the opening of the cavity were set large enough so that the laser pulse if able to fit completely inside the cavity.
For most simulations we used the cone angle $\tan\theta = 0.5$, depth and correspondingly the diameter of the opening equal to $\SI{30}{\um}$.

The wavelength of the laser is $\lambda = \SI{0.91}{\um}$, corresponding to the phase matching condition for the parametric amplification in DKDP crystal, which is believed to be the most viable option for generating laser radiation with the power well exceeding 10 PW mark \citep{XCELS-HPLSE, SEL2}.
The corresponding critical density for that wavelength is ${n_\mathrm{cr} \approx \SI{1.35e21}{\cm^{-3}}}$.
The numerical cell size is set to $0.015 \lambda \times 0.075\lambda\times 0.075\lambda$.
In the simulations the regular Yee scheme for solving Maxwell's equations was used with a time step ${c\Delta t = \Delta x / 4 = 0.00375\lambda}$.
According to the stability criterion of that scheme such cell size and the time step allow to simulate plasmas with the relativistically corrected density up to around $\num{3.8e3} n_\mathrm{cr}  \approx\SI{5e24}{\cm^{-3}}$, that is sufficient for the considered density of the target and also produced pair plasma.

\subsection*{Processing simulation results}

Averaged quantities in Table 1 are calculated using the data, measured in the $XY$ plane from a single layer of numerical cells, located in the middle of the simulation box ($z=0$), in the same way as presented in Fig.~\ref{fig-Density}. 
The averaging is done according to the following expression
\begin{equation}
  \langle A \rangle_w = \frac{\sum_{i,j} A_{i,j} w_{i,j} |y_j|}{\sum_{i,j} w_{i,j} |y_j|},
\end{equation}
where $i,j$ correspond to the cell index with the coordinates $(x_i, y_j)$ , $A_{i,j}$ the value of the quantity $A$ in that cell, $w_{i,j}$ is an arbitrary weight of the sell, and the additional factor $|y_j|\equiv r_\perp$ comes from the assumption that the distribution of the quantity $A$ has radial symmetry. 
For averaging the positron density we use the following weigh function
\begin{equation}
  w(x_i, y_j) = \begin{cases}
    1, \text{ if} \left( \frac{x_i - \overline{x}}{2 \sigma_x} \right)^2 + \left( \frac{y_j}{2 \sigma_y} \right)^2 < 1 \land n_{e^+} > n_\mathrm{cr}, \\
    0, \text{ otherwise}
  \end{cases},
\end{equation}
where $\overline{x} = \langle x \rangle_{n_{e^+}}$, ${\sigma_x^2 = \langle (x - \overline{x_i})^2 \rangle_{n_{e^+}}}$ and ${\sigma_y^2 = \langle y^2 \rangle_{n_{e^+}}}$ are the average $x$-coordinate and the standard deviations along the $x$ and $y$ directions of the positron density distribution correspondingly.
In other words, the positron density is averaged over a region within 2 standard deviations from the center of mass and densities smaller then $n_\mathrm{cr}$ are discarded from the calculations.

The quasi-static part of the magnetic field, plotted in Fig.~\ref{fig-Density}
is obtained by using a smooth low-pass filter with a cutoff frequency equal to $0.5 \times 2\pi/\lambda$.

The parameter $\nu$ is defined as a fraction of the positrons, contained in the volume of an ellipsoid with each semi-axis equal to twice the standard deviation of the positron density distribution, to the total amount of positrons, produced by the time of calculation.


\bibliography{bibliography, bib-extra}


\begin{thebibliography}{69}
\ifx \bisbn   \undefined \def \bisbn  #1{ISBN #1}\fi
\ifx \binits  \undefined \def \binits#1{#1}\fi
\ifx \bauthor  \undefined \def \bauthor#1{#1}\fi
\ifx \batitle  \undefined \def \batitle#1{#1}\fi
\ifx \bjtitle  \undefined \def \bjtitle#1{#1}\fi
\ifx \bvolume  \undefined \def \bvolume#1{\textbf{#1}}\fi
\ifx \byear  \undefined \def \byear#1{#1}\fi
\ifx \bissue  \undefined \def \bissue#1{#1}\fi
\ifx \bfpage  \undefined \def \bfpage#1{#1}\fi
\ifx \blpage  \undefined \def \blpage #1{#1}\fi
\ifx \burl  \undefined \def \burl#1{\textsf{#1}}\fi
\ifx \doiurl  \undefined \def \doiurl#1{\url{https://doi.org/#1}}\fi
\ifx \betal  \undefined \def \betal{\textit{et al.}}\fi
\ifx \binstitute  \undefined \def \binstitute#1{#1}\fi
\ifx \binstitutionaled  \undefined \def \binstitutionaled#1{#1}\fi
\ifx \bctitle  \undefined \def \bctitle#1{#1}\fi
\ifx \beditor  \undefined \def \beditor#1{#1}\fi
\ifx \bpublisher  \undefined \def \bpublisher#1{#1}\fi
\ifx \bbtitle  \undefined \def \bbtitle#1{#1}\fi
\ifx \bedition  \undefined \def \bedition#1{#1}\fi
\ifx \bseriesno  \undefined \def \bseriesno#1{#1}\fi
\ifx \blocation  \undefined \def \blocation#1{#1}\fi
\ifx \bsertitle  \undefined \def \bsertitle#1{#1}\fi
\ifx \bsnm \undefined \def \bsnm#1{#1}\fi
\ifx \bsuffix \undefined \def \bsuffix#1{#1}\fi
\ifx \bparticle \undefined \def \bparticle#1{#1}\fi
\ifx \barticle \undefined \def \barticle#1{#1}\fi
\bibcommenthead
\ifx \bconfdate \undefined \def \bconfdate #1{#1}\fi
\ifx \botherref \undefined \def \botherref #1{#1}\fi
\ifx \url \undefined \def \url#1{\textsf{#1}}\fi
\ifx \bchapter \undefined \def \bchapter#1{#1}\fi
\ifx \bbook \undefined \def \bbook#1{#1}\fi
\ifx \bcomment \undefined \def \bcomment#1{#1}\fi
\ifx \oauthor \undefined \def \oauthor#1{#1}\fi
\ifx \citeauthoryear \undefined \def \citeauthoryear#1{#1}\fi
\ifx \endbibitem  \undefined \def \endbibitem {}\fi
\ifx \bconflocation  \undefined \def \bconflocation#1{#1}\fi
\ifx \arxivurl  \undefined \def \arxivurl#1{\textsf{#1}}\fi
\csname PreBibitemsHook\endcsname

\bibitem[\protect\citeauthoryear{Tsai}{1974}]{tsaiPairProductionBremsstrahlung1974}
\begin{barticle}
\bauthor{\bsnm{Tsai}, \binits{Y.-S.}}:
\batitle{Pair production and bremsstrahlung of charged leptons}.
\bjtitle{Reviews of Modern Physics}
\bvolume{46}(\bissue{4}),
\bfpage{815}--\blpage{851}
(\byear{1974})
\doiurl{10.1103/RevModPhys.46.815}
\end{barticle}
\endbibitem

\bibitem[\protect\citeauthoryear{Bethe and Heitler}{1934}]{bethe1934stopping}
\begin{barticle}
\bauthor{\bsnm{Bethe}, \binits{H.}},
\bauthor{\bsnm{Heitler}, \binits{W.}}:
\batitle{On the stopping of fast particles and on the creation of positive electrons}.
\bjtitle{Proceedings of the Royal Society of London. Series A, Containing Papers of a Mathematical and Physical Character}
\bvolume{146}(\bissue{856}),
\bfpage{83}--\blpage{112}
(\byear{1934})
\doiurl{10.1098/rspa.1934.0140}
\end{barticle}
\endbibitem

\bibitem[\protect\citeauthoryear{Baier et~al.}{1998}]{Baier98}
\begin{bbook}
\bauthor{\bsnm{Baier}, \binits{V.N.}},
\bauthor{\bsnm{Katkov}, \binits{V.M.}},
\bauthor{\bsnm{Strakhovenko}, \binits{V.M.}}:
\bbtitle{Electromagnetic Processes at High Energies in Oriented Single Crystals}.
\bpublisher{World Scientific},
\blocation{Singapore}
(\byear{1998}).
\doiurl{10.1142/2216}
\end{bbook}
\endbibitem

\bibitem[\protect\citeauthoryear{Arrowsmith et~al.}{2024}]{arrowsmithLaboratoryRealizationRelativistic2024}
\begin{barticle}
\bauthor{\bsnm{Arrowsmith}, \binits{C.D.}},
\bauthor{\bsnm{Simon}, \binits{P.}},
\bauthor{\bsnm{Bilbao}, \binits{P.J.}},
\bauthor{\bsnm{Bott}, \binits{A.F.A.}},
\bauthor{\bsnm{Burger}, \binits{S.}},
\bauthor{\bsnm{Chen}, \binits{H.}},
\bauthor{\bsnm{Cruz}, \binits{F.D.}},
\bauthor{\bsnm{Davenne}, \binits{T.}},
\bauthor{\bsnm{Efthymiopoulos}, \binits{I.}},
\bauthor{\bsnm{Froula}, \binits{D.H.}},
\bauthor{\bsnm{Goillot}, \binits{A.}},
\bauthor{\bsnm{Gudmundsson}, \binits{J.T.}},
\bauthor{\bsnm{Haberberger}, \binits{D.}},
\bauthor{\bsnm{Halliday}, \binits{J.W.D.}},
\bauthor{\bsnm{Hodge}, \binits{T.}},
\bauthor{\bsnm{Huffman}, \binits{B.T.}},
\bauthor{\bsnm{Iaquinta}, \binits{S.}},
\bauthor{\bsnm{Miniati}, \binits{F.}},
\bauthor{\bsnm{Reville}, \binits{B.}},
\bauthor{\bsnm{Sarkar}, \binits{S.}},
\bauthor{\bsnm{Schekochihin}, \binits{A.A.}},
\bauthor{\bsnm{Silva}, \binits{L.O.}},
\bauthor{\bsnm{Simpson}, \binits{R.}},
\bauthor{\bsnm{Stergiou}, \binits{V.}},
\bauthor{\bsnm{Trines}, \binits{R.M.G.M.}},
\bauthor{\bsnm{Vieu}, \binits{T.}},
\bauthor{\bsnm{Charitonidis}, \binits{N.}},
\bauthor{\bsnm{Bingham}, \binits{R.}},
\bauthor{\bsnm{Gregori}, \binits{G.}}:
\batitle{Laboratory realization of relativistic pair-plasma beams}.
\bjtitle{Nature Communications}
\bvolume{15}(\bissue{1}),
\bfpage{5029}
(\byear{2024})
\doiurl{10.1038/s41467-024-49346-2}
\end{barticle}
\endbibitem

\bibitem[\protect\citeauthoryear{}{}]{ELI}
\begin{botherref}
The Extreme Light Infrastructure (ELI): \href{http://www.eli-laser.eu}{http://www.eli-laser.eu}
\end{botherref}
\endbibitem

\bibitem[\protect\citeauthoryear{Zou et~al.}{2015}]{Apollon}
\begin{barticle}
\bauthor{\bsnm{Zou}, \binits{{\relax JP}.}},
\bauthor{\bsnm{Le~Blanc}, \binits{C.}},
\bauthor{\bsnm{Papadopoulos}, \binits{{\relax DN}.}},
\bauthor{\bsnm{Ch{\'e}riaux}, \binits{G.}},
\bauthor{\bsnm{Georges}, \binits{P.}},
\bauthor{\bsnm{Mennerat}, \binits{G.}},
\bauthor{\bsnm{Druon}, \binits{F.}},
\bauthor{\bsnm{Lecherbourg}, \binits{L.}},
\bauthor{\bsnm{Pellegrina}, \binits{A.}},
\bauthor{\bsnm{Ramirez}, \binits{P.}}, \betal:
\batitle{Design and current progress of the {{Apollon}} 10 {{PW}} project}.
\bjtitle{High Power Laser Science and Engineering}
\bvolume{3},
\bfpage{2}
(\byear{2015})
\doiurl{10.1017/hpl.2014.41}
\end{barticle}
\endbibitem

\bibitem[\protect\citeauthoryear{Khazanov et~al.}{2023}]{XCELS-HPLSE}
\begin{botherref}
\oauthor{\bsnm{Khazanov}, \binits{E.}},
\oauthor{\bsnm{Shaykin}, \binits{A.}},
\oauthor{\bsnm{Kostyukov}, \binits{I.}},
\oauthor{\bsnm{Ginzburg}, \binits{V.}},
\oauthor{\bsnm{Mukhin}, \binits{I.}},
\oauthor{\bsnm{Yakovlev}, \binits{I.}},
\oauthor{\bsnm{Soloviev}, \binits{A.}},
\oauthor{\bsnm{Kuznetsov}, \binits{I.}},
\oauthor{\bsnm{Mironov}, \binits{S.}},
\oauthor{\bsnm{Korzhimanov}, \binits{A.}}, et al.:
Exawatt center for extreme light studies ({{XCELS}}).
High Power Laser Science and Engineering,
1--77
(2023)
\doiurl{10.1017/hpl.2023.69}
\end{botherref}
\endbibitem

\bibitem[\protect\citeauthoryear{Shao et~al.}{2020}]{SEL}
\begin{barticle}
\bauthor{\bsnm{Shao}, \binits{B.}},
\bauthor{\bsnm{Li}, \binits{Y.}},
\bauthor{\bsnm{Peng}, \binits{Y.}},
\bauthor{\bsnm{Wang}, \binits{P.}},
\bauthor{\bsnm{Qian}, \binits{J.}},
\bauthor{\bsnm{Leng}, \binits{Y.}},
\bauthor{\bsnm{Li}, \binits{R.}}:
\batitle{Broad-bandwidth high-temporal-contrast carrier-envelope-phase-stabilized laser seed for 100 {{PW}} lasers}.
\bjtitle{Optics Letters}
\bvolume{45}(\bissue{8}),
\bfpage{2215}--\blpage{2218}
(\byear{2020})
\doiurl{10.1364/OL.390110}
\end{barticle}
\endbibitem

\bibitem[\protect\citeauthoryear{Wu et~al.}{2022}]{SEL2}
\begin{barticle}
\bauthor{\bsnm{Wu}, \binits{F.}},
\bauthor{\bsnm{Hu}, \binits{J.}},
\bauthor{\bsnm{Liu}, \binits{X.}},
\bauthor{\bsnm{Zhang}, \binits{Z.}},
\bauthor{\bsnm{Bai}, \binits{P.}},
\bauthor{\bsnm{Wang}, \binits{X.}},
\bauthor{\bsnm{Zhao}, \binits{Y.}},
\bauthor{\bsnm{Yang}, \binits{X.}},
\bauthor{\bsnm{Xu}, \binits{Y.}},
\bauthor{\bsnm{Wang}, \binits{C.}},
\bauthor{\bsnm{Leng}, \binits{Y.}},
\bauthor{\bsnm{Li}, \binits{R.}}:
\batitle{Dispersion management for a 100 {{PW}} level laser using a mismatched-grating compressor}.
\bjtitle{High Power Laser Science and Engineering}
\bvolume{10},
\bfpage{38}
(\byear{2022})
\doiurl{10.1017/hpl.2022.29}
\end{barticle}
\endbibitem

\bibitem[\protect\citeauthoryear{Gan et~al.}{2021}]{SULF}
\begin{bchapter}
\bauthor{\bsnm{Gan}, \binits{Z.}},
\bauthor{\bsnm{Yu}, \binits{L.}},
\bauthor{\bsnm{Wang}, \binits{C.}},
\bauthor{\bsnm{Liu}, \binits{Y.}},
\bauthor{\bsnm{Xu}, \binits{Y.}},
\bauthor{\bsnm{Li}, \binits{W.}},
\bauthor{\bsnm{Li}, \binits{S.}},
\bauthor{\bsnm{Yu}, \binits{L.}},
\bauthor{\bsnm{Wang}, \binits{X.}},
\bauthor{\bsnm{Liu}, \binits{X.}},
\bauthor{\bsnm{Chen}, \binits{J.}},
\bauthor{\bsnm{Peng}, \binits{Y.}},
\bauthor{\bsnm{Xu}, \binits{L.}},
\bauthor{\bsnm{Yao}, \binits{B.}},
\bauthor{\bsnm{Zhang}, \binits{X.}},
\bauthor{\bsnm{Chen}, \binits{L.}},
\bauthor{\bsnm{Tang}, \binits{Y.}},
\bauthor{\bsnm{Wang}, \binits{X.}},
\bauthor{\bsnm{Yin}, \binits{D.}},
\bauthor{\bsnm{Liang}, \binits{X.}},
\bauthor{\bsnm{Leng}, \binits{Y.}},
\bauthor{\bsnm{Li}, \binits{R.}},
\bauthor{\bsnm{Xu}, \binits{Z.}}:
\bctitle{The {{Shanghai Superintense Ultrafast Laser Facility}} ({{SULF}}) {{Project}}}.
In: \beditor{\bsnm{Yamanouchi}, \binits{K.}},
\beditor{\bsnm{Midorikawa}, \binits{K.}},
\beditor{\bsnm{Roso}, \binits{L.}} (eds.)
\bbtitle{Progress in {{Ultrafast Intense Laser Science XVI}}}
vol. \bseriesno{141},
pp. \bfpage{199}--\blpage{217}.
\bpublisher{Springer},
\blocation{Cham}
(\byear{2021}).
\doiurl{10.1007/978-3-030-75089-3_10}
\end{bchapter}
\endbibitem

\bibitem[\protect\citeauthoryear{Veksler}{1957}]{veksler1957principle}
\begin{barticle}
\bauthor{\bsnm{Veksler}, \binits{V.I.}}:
\batitle{The principle of coherent acceleration of charged particles}.
\bjtitle{The Soviet Journal of Atomic Energy}
\bvolume{2}(\bissue{5}),
\bfpage{525}--\blpage{528}
(\byear{1957})
\doiurl{10.1007/BF01491001}
\end{barticle}
\endbibitem

\bibitem[\protect\citeauthoryear{Tajima and Dawson}{1979}]{tajima1979laser}
\begin{barticle}
\bauthor{\bsnm{Tajima}, \binits{T.}},
\bauthor{\bsnm{Dawson}, \binits{J.M.}}:
\batitle{Laser electron accelerator}.
\bjtitle{Physical Review Letters}
\bvolume{43}(\bissue{4}),
\bfpage{267}
(\byear{1979})
\doiurl{10.1103/PhysRevLett.43.267}
\end{barticle}
\endbibitem

\bibitem[\protect\citeauthoryear{Pukhov and {Meyer-ter-Vehn}}{2002}]{pukhovLaserWakeField2002}
\begin{barticle}
\bauthor{\bsnm{Pukhov}, \binits{A.}},
\bauthor{\bsnm{{Meyer-ter-Vehn}}, \binits{J.}}:
\batitle{Laser wake field acceleration: The highly non-linear broken-wave regime}.
\bjtitle{Applied Physics B: Photophysics and Laser Chemistry}
\bvolume{74}(\bissue{4}),
\bfpage{355}--\blpage{361}
(\byear{2002})
\doiurl{10.1007/s003400200795}
\end{barticle}
\endbibitem

\bibitem[\protect\citeauthoryear{Pukhov et~al.}{2004}]{faure2004laser}
\begin{barticle}
\bauthor{\bsnm{Pukhov}, \binits{A.}},
\bauthor{\bsnm{Kiselev}, \binits{S.}},
\bauthor{\bsnm{Gordienko}, \binits{S.}},
\bauthor{\bsnm{Lefebvre}, \binits{E.}},
\bauthor{\bsnm{Rousseau}, \binits{J.-P.}},
\bauthor{\bsnm{Burgy}, \binits{F.}}:
\batitle{A laser--plasma accelerator producing monoenergetic electron beams}.
\bjtitle{Nature}
\bvolume{431}(\bissue{7008}),
\bfpage{541}--\blpage{544}
(\byear{2004})
\doiurl{10.1038/nature02963}
\end{barticle}
\endbibitem

\bibitem[\protect\citeauthoryear{Esarey et~al.}{2009}]{esarey2009physics}
\begin{barticle}
\bauthor{\bsnm{Esarey}, \binits{E.}},
\bauthor{\bsnm{Schroeder}, \binits{C.B.}},
\bauthor{\bsnm{Leemans}, \binits{W.P.}}:
\batitle{Physics of laser-driven plasma-based electron accelerators}.
\bjtitle{Reviews of modern physics}
\bvolume{81}(\bissue{3}),
\bfpage{1229}
(\byear{2009})
\doiurl{10.1103/RevModPhys.81.1229}
\end{barticle}
\endbibitem

\bibitem[\protect\citeauthoryear{Clayton et~al.}{2010}]{clayton2010self}
\begin{barticle}
\bauthor{\bsnm{Clayton}, \binits{C.E.}},
\bauthor{\bsnm{Ralph}, \binits{{\relax JE}.}},
\bauthor{\bsnm{Albert}, \binits{F.}},
\bauthor{\bsnm{Fonseca}, \binits{{\relax RA}.}},
\bauthor{\bsnm{Glenzer}, \binits{{\relax SH}.}},
\bauthor{\bsnm{Joshi}, \binits{C.}},
\bauthor{\bsnm{Lu}, \binits{W.}},
\bauthor{\bsnm{Marsh}, \binits{{\relax KA}.}},
\bauthor{\bsnm{Martins}, \binits{S.F.}},
\bauthor{\bsnm{Mori}, \binits{W.B.}}, \betal:
\batitle{Self-guided laser wakefield acceleration beyond 1 {{GeV}} using ionization-induced injection}.
\bjtitle{Physical Review Letters}
\bvolume{105}(\bissue{10}),
\bfpage{105003}
(\byear{2010})
\doiurl{10.1103/PhysRevLett.105.105003}
\end{barticle}
\endbibitem

\bibitem[\protect\citeauthoryear{Kostyukov and Pukhov}{2015}]{Kostyukov2015UFN}
\begin{barticle}
\bauthor{\bsnm{Kostyukov}, \binits{I.}},
\bauthor{\bsnm{Pukhov}, \binits{A.M.}}:
\batitle{Plasma-based methods for electron acceleration: Current status and prospects}.
\bjtitle{Physics-Uspekhi}
\bvolume{58}(\bissue{1}),
\bfpage{81}
(\byear{2015})
\doiurl{10.3367/UFNe.0185.201501g.0089}
\end{barticle}
\endbibitem

\bibitem[\protect\citeauthoryear{Wen et~al.}{2019}]{wenPolarizedLaserwakefieldacceleratedKiloampere2019}
\begin{barticle}
\bauthor{\bsnm{Wen}, \binits{M.}},
\bauthor{\bsnm{Tamburini}, \binits{M.}},
\bauthor{\bsnm{Keitel}, \binits{C.H.}}:
\batitle{Polarized laser-wakefield-accelerated kiloampere electron beams}.
\bjtitle{Physical Review Letters}
\bvolume{122}(\bissue{21}),
\bfpage{214801}
(\byear{2019})
\doiurl{10.1103/physrevlett.122.214801}
\end{barticle}
\endbibitem

\bibitem[\protect\citeauthoryear{Palastro et~al.}{2020}]{palastro2020dephasingless}
\begin{barticle}
\bauthor{\bsnm{Palastro}, \binits{{\relax JP}.}},
\bauthor{\bsnm{Shaw}, \binits{{\relax JL}.}},
\bauthor{\bsnm{Franke}, \binits{P.}},
\bauthor{\bsnm{Ramsey}, \binits{D.}},
\bauthor{\bsnm{Simpson}, \binits{{\relax TT}.}},
\bauthor{\bsnm{Froula}, \binits{{\relax DH}.}}:
\batitle{Dephasingless laser wakefield acceleration}.
\bjtitle{Physical Review Letters}
\bvolume{124}(\bissue{13}),
\bfpage{134802}
(\byear{2020})
\doiurl{10.1103/PhysRevLett.124.134802}
\end{barticle}
\endbibitem

\bibitem[\protect\citeauthoryear{Tajima et~al.}{2020}]{tajima2020wakefield}
\begin{barticle}
\bauthor{\bsnm{Tajima}, \binits{T.}},
\bauthor{\bsnm{Yan}, \binits{{\relax XQ}.}},
\bauthor{\bsnm{Ebisuzaki}, \binits{T.}}:
\batitle{Wakefield acceleration}.
\bjtitle{Reviews of Modern Plasma Physics}
\bvolume{4},
\bfpage{1}--\blpage{72}
(\byear{2020})
\doiurl{10.1007/s41614-020-0043-z}
\end{barticle}
\endbibitem

\bibitem[\protect\citeauthoryear{Sarri et~al.}{2015a}]{sarriGenerationNeutralHighdensity2015}
\begin{barticle}
\bauthor{\bsnm{Sarri}, \binits{G.}},
\bauthor{\bsnm{Poder}, \binits{K.}},
\bauthor{\bsnm{Cole}, \binits{J.M.}},
\bauthor{\bsnm{Schumaker}, \binits{W.}},
\bauthor{\bsnm{Di~Piazza}, \binits{A.}},
\bauthor{\bsnm{Reville}, \binits{B.}},
\bauthor{\bsnm{Dzelzainis}, \binits{T.}},
\bauthor{\bsnm{Doria}, \binits{D.}},
\bauthor{\bsnm{Gizzi}, \binits{L.A.}},
\bauthor{\bsnm{Grittani}, \binits{G.}},
\bauthor{\bsnm{Kar}, \binits{S.}},
\bauthor{\bsnm{Keitel}, \binits{C.H.}},
\bauthor{\bsnm{Krushelnick}, \binits{K.}},
\bauthor{\bsnm{Kuschel}, \binits{S.}},
\bauthor{\bsnm{Mangles}, \binits{S.P.D.}},
\bauthor{\bsnm{Najmudin}, \binits{Z.}},
\bauthor{\bsnm{Shukla}, \binits{N.}},
\bauthor{\bsnm{Silva}, \binits{L.O.}},
\bauthor{\bsnm{Symes}, \binits{D.}},
\bauthor{\bsnm{Thomas}, \binits{A.G.R.}},
\bauthor{\bsnm{Vargas}, \binits{M.}},
\bauthor{\bsnm{Vieira}, \binits{J.}},
\bauthor{\bsnm{Zepf}, \binits{M.}}:
\batitle{Generation of neutral and high-density electron--positron pair plasmas in the laboratory}.
\bjtitle{Nature Communications}
\bvolume{6}(\bissue{1}),
\bfpage{6747}
(\byear{2015})
\doiurl{10.1038/ncomms7747}
\end{barticle}
\endbibitem

\bibitem[\protect\citeauthoryear{Sarri et~al.}{2015b}]{sarriOverviewLaserdrivenGeneration2015}
\begin{barticle}
\bauthor{\bsnm{Sarri}, \binits{G.}},
\bauthor{\bsnm{Dieckmann}, \binits{M.E.}},
\bauthor{\bsnm{Kourakis}, \binits{I.}},
\bauthor{\bsnm{Di~Piazza}, \binits{A.}},
\bauthor{\bsnm{Reville}, \binits{B.}},
\bauthor{\bsnm{Keitel}, \binits{C.H.}},
\bauthor{\bsnm{Zepf}, \binits{M.}}:
\batitle{Overview of laser-driven generation of electron--positron beams}.
\bjtitle{Journal of Plasma Physics}
\bvolume{81}(\bissue{4}),
\bfpage{455810401}
(\byear{2015})
\doiurl{10.1017/S002237781500046X}
\end{barticle}
\endbibitem

\bibitem[\protect\citeauthoryear{Berestetskii et~al.}{1982}]{Berestetskii82}
\begin{bbook}
\bauthor{\bsnm{Berestetskii}, \binits{V.B.}},
\bauthor{\bsnm{Lifshitz}, \binits{E.M.}},
\bauthor{\bsnm{Pitaevskii}, \binits{L.P.}}:
\bbtitle{Quantum Electrodynamics}.
\bpublisher{Pergamon},
\blocation{New York}
(\byear{1982}).
\doiurl{10.1016/C2009-0-24486-2}
\end{bbook}
\endbibitem

\bibitem[\protect\citeauthoryear{Elkina et~al.}{2011}]{elkinaQEDCascadesInduced2011}
\begin{barticle}
\bauthor{\bsnm{Elkina}, \binits{N.V.}},
\bauthor{\bsnm{Fedotov}, \binits{A.M.}},
\bauthor{\bsnm{Kostyukov}, \binits{I.Y.}},
\bauthor{\bsnm{Legkov}, \binits{M.V.}},
\bauthor{\bsnm{Narozhny}, \binits{N.B.}},
\bauthor{\bsnm{Nerush}, \binits{E.N.}},
\bauthor{\bsnm{Ruhl}, \binits{H.}}:
\batitle{{{QED}} cascades induced by circularly polarized laser fields}.
\bjtitle{Physical Review Special Topics - Accelerators and Beams}
\bvolume{14}(\bissue{5}),
\bfpage{054401}
(\byear{2011})
\doiurl{10.1103/PhysRevSTAB.14.054401}
{\href{https://arxiv.org/abs/1010.4528}{{1010.4528}}}
\end{barticle}
\endbibitem

\bibitem[\protect\citeauthoryear{Nerush and Kostyukov}{2007}]{nerushRadiationEmissionExtreme2007}
\begin{barticle}
\bauthor{\bsnm{Nerush}, \binits{E.}},
\bauthor{\bsnm{Kostyukov}, \binits{I.}}:
\batitle{Radiation {{Emission}} by {{Extreme Relativistic Electrons}} and {{Pair Production}} by {{Hard Photons}} in a {{Strong Plasma Wakefield}}}.
\bjtitle{Physical Review E}
\bvolume{75}(\bissue{5}),
\bfpage{057401}
(\byear{2007})
\doiurl{10.1103/PhysRevE.75.057401}
\end{barticle}
\endbibitem

\bibitem[\protect\citeauthoryear{Bell and Kirk}{2008}]{bellPossibilityProlificPair2008}
\begin{barticle}
\bauthor{\bsnm{Bell}, \binits{A.R.}},
\bauthor{\bsnm{Kirk}, \binits{J.G.}}:
\batitle{Possibility of {{Prolific Pair Production}} with {{High-Power Lasers}}}.
\bjtitle{Physical Review Letters}
\bvolume{101}(\bissue{20}),
\bfpage{200403}
(\byear{2008})
\doiurl{10.1103/PhysRevLett.101.200403}
\end{barticle}
\endbibitem

\bibitem[\protect\citeauthoryear{Nerush et~al.}{2011}]{nerushLaserFieldAbsorption2011}
\begin{barticle}
\bauthor{\bsnm{Nerush}, \binits{E.N.}},
\bauthor{\bsnm{Kostyukov}, \binits{I.{\relax Yu}.}},
\bauthor{\bsnm{Fedotov}, \binits{A.M.}},
\bauthor{\bsnm{Narozhny}, \binits{N.B.}},
\bauthor{\bsnm{Elkina}, \binits{N.V.}},
\bauthor{\bsnm{Ruhl}, \binits{H.}}:
\batitle{Laser {{Field Absorption}} in {{Self-Generated Electron-Positron Pair Plasma}}}.
\bjtitle{Physical Review Letters}
\bvolume{106}(\bissue{3}),
\bfpage{035001}
(\byear{2011})
\doiurl{10.1103/PhysRevLett.106.035001}
\end{barticle}
\endbibitem

\bibitem[\protect\citeauthoryear{Ridgers et~al.}{2012}]{ridgersDenseElectronPositronPlasmas2012}
\begin{barticle}
\bauthor{\bsnm{Ridgers}, \binits{C.P.}},
\bauthor{\bsnm{Brady}, \binits{C.S.}},
\bauthor{\bsnm{Duclous}, \binits{R.}},
\bauthor{\bsnm{Kirk}, \binits{J.G.}},
\bauthor{\bsnm{Bennett}, \binits{K.}},
\bauthor{\bsnm{Arber}, \binits{T.D.}},
\bauthor{\bsnm{Robinson}, \binits{A.P.L.}},
\bauthor{\bsnm{Bell}, \binits{A.R.}}:
\batitle{Dense {{Electron-Positron Plasmas}} and {{Ultra-Intense Bursts}} of {{Gamma-Rays}} from {{Laser-Irradiated Solids}}}.
\bjtitle{Physical Review Letters}
\bvolume{108}(\bissue{16}),
\bfpage{165006}
(\byear{2012})
\doiurl{10.1103/PhysRevLett.108.165006}
{[physics]}
\end{barticle}
\endbibitem

\bibitem[\protect\citeauthoryear{Kirk et~al.}{2013}]{kirkPairPlasmaCushions2013}
\begin{barticle}
\bauthor{\bsnm{Kirk}, \binits{J.G.}},
\bauthor{\bsnm{Bell}, \binits{A.R.}},
\bauthor{\bsnm{Ridgers}, \binits{C.P.}}:
\batitle{Pair plasma cushions in the hole-boring scenario}.
\bjtitle{Plasma Physics and Controlled Fusion}
\bvolume{55}(\bissue{9}),
\bfpage{095016}
(\byear{2013})
\doiurl{10.1088/0741-3335/55/9/095016}
\end{barticle}
\endbibitem

\bibitem[\protect\citeauthoryear{Narozhny and Fedotov}{2015}]{narozhnyQuantumelectrodynamicCascadesIntense2015}
\begin{barticle}
\bauthor{\bsnm{Narozhny}, \binits{N.B.}},
\bauthor{\bsnm{Fedotov}, \binits{A.M.}}:
\batitle{Quantum-electrodynamic cascades in intense laser fields}.
\bjtitle{Physics-Uspekhi}
\bvolume{58}(\bissue{1}),
\bfpage{95}
(\byear{2015})
\doiurl{10.3367/UFNe.0185.201501i.0103}
\end{barticle}
\endbibitem

\bibitem[\protect\citeauthoryear{Zhu et~al.}{2016}]{zhuDenseGeVElectron2016}
\begin{barticle}
\bauthor{\bsnm{Zhu}, \binits{X.-L.}},
\bauthor{\bsnm{Yu}, \binits{T.-P.}},
\bauthor{\bsnm{Sheng}, \binits{Z.-M.}},
\bauthor{\bsnm{Yin}, \binits{Y.}},
\bauthor{\bsnm{Turcu}, \binits{I.C.E.}},
\bauthor{\bsnm{Pukhov}, \binits{A.}}:
\batitle{Dense {{GeV}} electron--positron pairs generated by lasers in near-critical-density plasmas}.
\bjtitle{Nature Communications}
\bvolume{7}(\bissue{1}),
\bfpage{13686}
(\byear{2016})
\doiurl{10.1038/ncomms13686}
\end{barticle}
\endbibitem

\bibitem[\protect\citeauthoryear{Kostyukov and Nerush}{2016}]{kostyukovProductionDynamicsPositrons2016}
\begin{barticle}
\bauthor{\bsnm{Kostyukov}, \binits{I.{\relax Yu}.}},
\bauthor{\bsnm{Nerush}, \binits{E.N.}}:
\batitle{Production and dynamics of positrons in ultrahigh intensity laser-foil interactions}.
\bjtitle{Physics of Plasmas}
\bvolume{23}(\bissue{9}),
\bfpage{093119}
(\byear{2016})
\doiurl{10.1063/1.4962567}
\end{barticle}
\endbibitem

\bibitem[\protect\citeauthoryear{Grismayer et~al.}{2017}]{grismayerSeededQEDCascades2017}
\begin{barticle}
\bauthor{\bsnm{Grismayer}, \binits{T.}},
\bauthor{\bsnm{Vranic}, \binits{M.}},
\bauthor{\bsnm{Martins}, \binits{J.L.}},
\bauthor{\bsnm{Fonseca}, \binits{R.A.}},
\bauthor{\bsnm{Silva}, \binits{L.O.}}:
\batitle{Seeded {{QED}} cascades in counterpropagating laser pulses}.
\bjtitle{Physical Review E}
\bvolume{95}(\bissue{2}),
\bfpage{023210}
(\byear{2017})
\doiurl{10.1103/PhysRevE.95.023210}
\end{barticle}
\endbibitem

\bibitem[\protect\citeauthoryear{Jirka et~al.}{2017}]{jirkaQEDCascade102017}
\begin{barticle}
\bauthor{\bsnm{Jirka}, \binits{M.}},
\bauthor{\bsnm{Klimo}, \binits{O.}},
\bauthor{\bsnm{Vranic}, \binits{M.}},
\bauthor{\bsnm{Weber}, \binits{S.}},
\bauthor{\bsnm{Korn}, \binits{G.}}:
\batitle{{{QED}} cascade with 10 {{PW-class}} lasers}.
\bjtitle{Scientific Reports}
\bvolume{7}(\bissue{1}),
\bfpage{15302}
(\byear{2017})
\doiurl{10.1038/s41598-017-15747-1}
\end{barticle}
\endbibitem

\bibitem[\protect\citeauthoryear{Luo et~al.}{2018}]{luoQEDCascadeSaturation2018}
\begin{barticle}
\bauthor{\bsnm{Luo}, \binits{W.}},
\bauthor{\bsnm{Liu}, \binits{W.-Y.}},
\bauthor{\bsnm{Yuan}, \binits{T.}},
\bauthor{\bsnm{Chen}, \binits{M.}},
\bauthor{\bsnm{Yu}, \binits{J.-Y.}},
\bauthor{\bsnm{Li}, \binits{F.-Y.}},
\bauthor{\bsnm{Del~Sorbo}, \binits{D.}},
\bauthor{\bsnm{Ridgers}, \binits{C.P.}},
\bauthor{\bsnm{Sheng}, \binits{Z.-M.}}:
\batitle{{{QED}} cascade saturation in extreme high fields}.
\bjtitle{Scientific Reports}
\bvolume{8}(\bissue{1}),
\bfpage{8400}
(\byear{2018})
\doiurl{10.1038/s41598-018-26785-8}
\end{barticle}
\endbibitem

\bibitem[\protect\citeauthoryear{Yuan et~al.}{2018}]{yuanSpatiotemporalDistributionsPair2018}
\begin{barticle}
\bauthor{\bsnm{Yuan}, \binits{T.}},
\bauthor{\bsnm{Yu}, \binits{J.Y.}},
\bauthor{\bsnm{Liu}, \binits{W.Y.}},
\bauthor{\bsnm{Weng}, \binits{S.M.}},
\bauthor{\bsnm{Yuan}, \binits{X.H.}},
\bauthor{\bsnm{Luo}, \binits{W.}},
\bauthor{\bsnm{Chen}, \binits{M.}},
\bauthor{\bsnm{Sheng}, \binits{Z.M.}},
\bauthor{\bsnm{Zhang}, \binits{J.}}:
\batitle{Spatiotemporal distributions of pair production and cascade in solid targets irradiated by ultra-relativistic lasers with different polarizations}.
\bjtitle{Plasma Physics and Controlled Fusion}
\bvolume{60}(\bissue{6}),
\bfpage{065003}
(\byear{2018})
\doiurl{10.1088/1361-6587/aab3ba}
\end{barticle}
\endbibitem

\bibitem[\protect\citeauthoryear{Sorbo et~al.}{2018}]{sorboEfficientIonAcceleration2018}
\begin{barticle}
\bauthor{\bsnm{Sorbo}, \binits{D.D.}},
\bauthor{\bsnm{Blackman}, \binits{D.R.}},
\bauthor{\bsnm{Capdessus}, \binits{R.}},
\bauthor{\bsnm{Small}, \binits{K.}},
\bauthor{\bsnm{{Slade-Lowther}}, \binits{C.}},
\bauthor{\bsnm{Luo}, \binits{W.}},
\bauthor{\bsnm{Duff}, \binits{M.J.}},
\bauthor{\bsnm{Robinson}, \binits{A.P.L.}},
\bauthor{\bsnm{McKenna}, \binits{P.}},
\bauthor{\bsnm{Sheng}, \binits{Z.-M.}},
\bauthor{\bsnm{Pasley}, \binits{J.}},
\bauthor{\bsnm{Ridgers}, \binits{C.P.}}:
\batitle{Efficient ion acceleration and dense electron--positron plasma creation in ultra-high intensity laser-solid interactions}.
\bjtitle{New Journal of Physics}
\bvolume{20}(\bissue{3}),
\bfpage{033014}
(\byear{2018})
\doiurl{10.1088/1367-2630/aaae61}
\end{barticle}
\endbibitem

\bibitem[\protect\citeauthoryear{Lu et~al.}{2018}]{luEnhancedCopiousElectron2018}
\begin{barticle}
\bauthor{\bsnm{Lu}, \binits{Y.}},
\bauthor{\bsnm{Yu}, \binits{T.-P.}},
\bauthor{\bsnm{Hu}, \binits{L.-X.}},
\bauthor{\bsnm{Ge}, \binits{Z.-Y.}},
\bauthor{\bsnm{Wang}, \binits{W.-Q.}},
\bauthor{\bsnm{Liu}, \binits{J.-X.}},
\bauthor{\bsnm{Liu}, \binits{K.}},
\bauthor{\bsnm{Yin}, \binits{Y.}},
\bauthor{\bsnm{Shao}, \binits{F.-Q.}}:
\batitle{Enhanced copious electron--positron pair production via electron injection from a mass-limited foil}.
\bjtitle{Plasma Physics and Controlled Fusion}
\bvolume{60}(\bissue{12}),
\bfpage{125008}
(\byear{2018})
\doiurl{10.1088/1361-6587/aae819}
\end{barticle}
\endbibitem

\bibitem[\protect\citeauthoryear{Samsonov et~al.}{2019}]{samsonov2019laser}
\begin{barticle}
\bauthor{\bsnm{Samsonov}, \binits{A.S.}},
\bauthor{\bsnm{Nerush}, \binits{E.N.}},
\bauthor{\bsnm{Kostyukov}, \binits{I.}}:
\batitle{Laser-driven vacuum breakdown waves}.
\bjtitle{Scientific reports}
\bvolume{9}(\bissue{1}),
\bfpage{11133}
(\byear{2019})
\doiurl{10.1038/s41598-019-47355-6}
\end{barticle}
\endbibitem

\bibitem[\protect\citeauthoryear{Efimenko et~al.}{2019}]{efimenkoLaserdrivenPlasmaPinching2019}
\begin{barticle}
\bauthor{\bsnm{Efimenko}, \binits{E.S.}},
\bauthor{\bsnm{Bashinov}, \binits{A.V.}},
\bauthor{\bsnm{Gonoskov}, \binits{A.A.}},
\bauthor{\bsnm{Bastrakov}, \binits{S.I.}},
\bauthor{\bsnm{Muraviev}, \binits{A.A.}},
\bauthor{\bsnm{Meyerov}, \binits{I.B.}},
\bauthor{\bsnm{Kim}, \binits{A.V.}},
\bauthor{\bsnm{Sergeev}, \binits{A.M.}}:
\batitle{Laser-driven plasma pinching in $e^-e^+$ cascade}.
\bjtitle{Physical Review E}
\bvolume{99}(\bissue{3}),
\bfpage{031201}
(\byear{2019})
\doiurl{10.1103/PhysRevE.99.031201}
\end{barticle}
\endbibitem

\bibitem[\protect\citeauthoryear{Chen and Fiuza}{2023}]{chenPerspectivesRelativisticElectron2023}
\begin{barticle}
\bauthor{\bsnm{Chen}, \binits{H.}},
\bauthor{\bsnm{Fiuza}, \binits{F.}}:
\batitle{Perspectives on relativistic electron--positron pair plasma experiments of astrophysical relevance using high-power lasers}.
\bjtitle{Physics of Plasmas}
\bvolume{30}(\bissue{2}),
\bfpage{020601}
(\byear{2023})
\doiurl{10.1063/5.0134819}
\end{barticle}
\endbibitem

\bibitem[\protect\citeauthoryear{Serebryakov et~al.}{2017}]{serebryakovNearsurfaceElectronAcceleration2017}
\begin{barticle}
\bauthor{\bsnm{Serebryakov}, \binits{D.A.}},
\bauthor{\bsnm{Nerush}, \binits{E.N.}},
\bauthor{\bsnm{Kostyukov}, \binits{I.{\relax Yu}.}}:
\batitle{Near-surface electron acceleration during intense laser--solid interaction in the grazing incidence regime}.
\bjtitle{Physics of Plasmas}
\bvolume{24}(\bissue{12}),
\bfpage{123115}
(\byear{2017})
\doiurl{10.1063/1.5002671}
\end{barticle}
\endbibitem

\bibitem[\protect\citeauthoryear{Shen et~al.}{2024}]{shen2024high}
\begin{barticle}
\bauthor{\bsnm{Shen}, \binits{X.}},
\bauthor{\bsnm{Pukhov}, \binits{A.}},
\bauthor{\bsnm{Qiao}, \binits{B.}}:
\batitle{High-flux bright x-ray source from femtosecond laser-irradiated microtapes}.
\bjtitle{Communications Physics}
\bvolume{7}(\bissue{1}),
\bfpage{84}
(\byear{2024})
\doiurl{10.1038/s42005-024-01575-z}
\end{barticle}
\endbibitem

\bibitem[\protect\citeauthoryear{Filipovic and Pukhov}{2022}]{filipovicQEDEffectsGrazing2022}
\begin{barticle}
\bauthor{\bsnm{Filipovic}, \binits{M.}},
\bauthor{\bsnm{Pukhov}, \binits{A.}}:
\batitle{{{QED Effects}} at {{Grazing Incidence}} on {{Solid-State-Targets}}}.
\bjtitle{The European Physical Journal D}
\bvolume{76}(\bissue{10}),
\bfpage{187}
(\byear{2022})
\doiurl{10.1140/epjd/s10053-022-00494-4}
\end{barticle}
\endbibitem

\bibitem[\protect\citeauthoryear{Samsonov et~al.}{2023}]{samsonovGenerationElectronPositron2023}
\begin{barticle}
\bauthor{\bsnm{Samsonov}, \binits{A.S.}},
\bauthor{\bsnm{Kostyukov}, \binits{I.{\relax Yu}.}},
\bauthor{\bsnm{Filipovic}, \binits{M.}},
\bauthor{\bsnm{Pukhov}, \binits{A.M.}}:
\batitle{Generation of {{Electron}}--{{Positron Pairs}} upon {{Grazing Incidence}} of a {{Laser Pulse}} on a {{Foil}}}.
\bjtitle{Bulletin of the Lebedev Physics Institute}
\bvolume{50}(\bissue{S6}),
\bfpage{693}--\blpage{699}
(\byear{2023})
\doiurl{10.3103/S1068335623180112}
\end{barticle}
\endbibitem

\bibitem[\protect\citeauthoryear{Pitaevskii}{1961}]{pitaevskiiElectricForcesTransparent1961}
\begin{barticle}
\bauthor{\bsnm{Pitaevskii}, \binits{L.}}:
\batitle{Electric {{Forces}} in a {{Transparent Dispersive Medium}}}.
\bjtitle{SOVIET PHYSICS JETP-USSR}
\bvolume{12}(\bissue{5}),
\bfpage{1008}--\blpage{1013}
(\byear{1961})
\end{barticle}
\endbibitem

\bibitem[\protect\citeauthoryear{Sheng and {Meyer-ter-Vehn}}{1996}]{shengInverseFaradayEffect1996}
\begin{barticle}
\bauthor{\bsnm{Sheng}, \binits{Z.M.}},
\bauthor{\bsnm{{Meyer-ter-Vehn}}, \binits{J.}}:
\batitle{Inverse {{Faraday}} effect and propagation of circularly polarized intense laser beams in plasmas}.
\bjtitle{Physical Review E}
\bvolume{54}(\bissue{2}),
\bfpage{1833}--\blpage{1842}
(\byear{1996})
\doiurl{10.1103/PhysRevE.54.1833}
\end{barticle}
\endbibitem

\bibitem[\protect\citeauthoryear{Haines}{2001}]{hainesGenerationAxialMagnetic2001}
\begin{barticle}
\bauthor{\bsnm{Haines}, \binits{M.G.}}:
\batitle{Generation of an {{Axial Magnetic Field}} from {{Photon Spin}}}.
\bjtitle{Physical Review Letters}
\bvolume{87}(\bissue{13}),
\bfpage{135005}
(\byear{2001})
\doiurl{10.1103/PhysRevLett.87.135005}
\end{barticle}
\endbibitem

\bibitem[\protect\citeauthoryear{Shvets et~al.}{2002}]{shvetsMagneticFieldGeneration2002}
\begin{barticle}
\bauthor{\bsnm{Shvets}, \binits{G.}},
\bauthor{\bsnm{Fisch}, \binits{N.J.}},
\bauthor{\bsnm{Rax}, \binits{J.-M.}}:
\batitle{Magnetic field generation through angular momentum exchange between circularly polarized radiation and charged particles}.
\bjtitle{Physical Review E}
\bvolume{65}(\bissue{4}),
\bfpage{046403}
(\byear{2002})
\doiurl{10.1103/PhysRevE.65.046403}
\end{barticle}
\endbibitem

\bibitem[\protect\citeauthoryear{Kostyukov et~al.}{2002}]{kostyukovMagneticfieldGenerationElectron2002}
\begin{barticle}
\bauthor{\bsnm{Kostyukov}, \binits{I.{\relax Yu}.}},
\bauthor{\bsnm{Shvets}, \binits{G.}},
\bauthor{\bsnm{Fisch}, \binits{N.J.}},
\bauthor{\bsnm{Rax}, \binits{J.M.}}:
\batitle{Magnetic-field generation and electron acceleration in relativistic laser channel}.
\bjtitle{Physics of Plasmas}
\bvolume{9}(\bissue{2}),
\bfpage{636}--\blpage{648}
(\byear{2002})
\doiurl{10.1063/1.1430436}
\end{barticle}
\endbibitem

\bibitem[\protect\citeauthoryear{Nuter et~al.}{2020}]{nuterGainElectronOrbital2020}
\begin{barticle}
\bauthor{\bsnm{Nuter}, \binits{R.}},
\bauthor{\bsnm{Korneev}, \binits{{\relax Ph}.}},
\bauthor{\bsnm{Dmitriev}, \binits{E.}},
\bauthor{\bsnm{Thiele}, \binits{I.}},
\bauthor{\bsnm{Tikhonchuk}, \binits{V.T.}}:
\batitle{Gain of electron orbital angular momentum in a direct laser acceleration process}.
\bjtitle{Physical Review E}
\bvolume{101}(\bissue{5}),
\bfpage{053202}
(\byear{2020})
\doiurl{10.1103/PhysRevE.101.053202}
\end{barticle}
\endbibitem

\bibitem[\protect\citeauthoryear{Jiang et~al.}{2021}]{jiangMagneticFieldAmplification2021}
\begin{barticle}
\bauthor{\bsnm{Jiang}, \binits{K.}},
\bauthor{\bsnm{Pukhov}, \binits{A.}},
\bauthor{\bsnm{Zhou}, \binits{C.T.}}:
\batitle{Magnetic field amplification to gigagauss scale via hydrodynamic flows and dynamos driven by femtosecond lasers}.
\bjtitle{New Journal of Physics}
\bvolume{23}(\bissue{6}),
\bfpage{063054}
(\byear{2021})
\doiurl{10.1088/1367-2630/ac0573}
\end{barticle}
\endbibitem

\bibitem[\protect\citeauthoryear{Longman and Fedosejevs}{2021}]{longmanKiloTeslaAxialMagnetic2021}
\begin{barticle}
\bauthor{\bsnm{Longman}, \binits{A.}},
\bauthor{\bsnm{Fedosejevs}, \binits{R.}}:
\batitle{Kilo-{{Tesla}} axial magnetic field generation with high intensity spin and orbital angular momentum beams}.
\bjtitle{Physical Review Research}
\bvolume{3}(\bissue{4}),
\bfpage{043180}
(\byear{2021})
\doiurl{10.1103/PhysRevResearch.3.043180}
\end{barticle}
\endbibitem

\bibitem[\protect\citeauthoryear{Liseykina et~al.}{2016}]{liseykinaInverseFaradayEffect2016}
\begin{barticle}
\bauthor{\bsnm{Liseykina}, \binits{T.V.}},
\bauthor{\bsnm{Popruzhenko}, \binits{S.V.}},
\bauthor{\bsnm{Macchi}, \binits{A.}}:
\batitle{Inverse {{Faraday}} effect driven by radiation friction}.
\bjtitle{New Journal of Physics}
\bvolume{18}(\bissue{7}),
\bfpage{072001}
(\byear{2016})
\doiurl{10.1088/1367-2630/18/7/072001}
\end{barticle}
\endbibitem

\bibitem[\protect\citeauthoryear{Liseykina et~al.}{2021}]{liseykinaQuantumEffectsRadiation2021}
\begin{barticle}
\bauthor{\bsnm{Liseykina}, \binits{T.V.}},
\bauthor{\bsnm{Macchi}, \binits{A.}},
\bauthor{\bsnm{Popruzhenko}, \binits{S.V.}}:
\batitle{Quantum effects on radiation friction driven magnetic field generation}.
\bjtitle{The European Physical Journal Plus}
\bvolume{136}(\bissue{2}),
\bfpage{170}
(\byear{2021})
\doiurl{10.1140/epjp/s13360-020-01030-2}
\end{barticle}
\endbibitem

\bibitem[\protect\citeauthoryear{Liseykina et~al.}{2023}]{liseykina2023inverse}
\begin{barticle}
\bauthor{\bsnm{Liseykina}, \binits{{\relax TV}.}},
\bauthor{\bsnm{Peganov}, \binits{{\relax EE}.}},
\bauthor{\bsnm{Popruzhenko}, \binits{{\relax SV}.}}:
\batitle{The inverse faraday effect induced by radiation friction during irradiation of dense plasma with crossed multipetawatt laser beams}.
\bjtitle{Bulletin of the Lebedev Physics Institute}
\bvolume{50}(\bissue{Suppl 6}),
\bfpage{700}--\blpage{705}
(\byear{2023})
\doiurl{10.3103/S1068335623180082}
\end{barticle}
\endbibitem

\bibitem[\protect\citeauthoryear{Samsonov et~al.}{2021}]{samsonovEffectElectronPositron2021}
\begin{barticle}
\bauthor{\bsnm{Samsonov}, \binits{A.S.}},
\bauthor{\bsnm{Nerush}, \binits{E.N.}},
\bauthor{\bsnm{Kostyukov}, \binits{{\relax I. Yu}.}}:
\batitle{Effect of electron--positron plasma production on the generation of a magnetic field in laser-plasma interactions}.
\bjtitle{Quantum Electronics}
\bvolume{51}(\bissue{10}),
\bfpage{861}--\blpage{865}
(\byear{2021})
\doiurl{10.1070/QEL17601}
\end{barticle}
\endbibitem

\bibitem[\protect\citeauthoryear{De~Groot}{1980}]{degrootRelativisticKineticTheory1980}
\begin{bbook}
\bauthor{\bsnm{De~Groot}, \binits{S.R.}}:
\bbtitle{Relativistic {{Kinetic Theory}}. {{Principles}} and {{Applications}}},
(\byear{1980})
\end{bbook}
\endbibitem

\bibitem[\protect\citeauthoryear{Helander and Connor}{2016}]{helanderGyrokineticStabilityTheory2016}
\begin{barticle}
\bauthor{\bsnm{Helander}, \binits{P.}},
\bauthor{\bsnm{Connor}, \binits{J.W.}}:
\batitle{Gyrokinetic stability theory of electron--positron plasmas}.
\bjtitle{Journal of Plasma Physics}
\bvolume{82}(\bissue{3}),
\bfpage{905820301}
(\byear{2016})
\doiurl{10.1017/S0022377816000490}
\end{barticle}
\endbibitem

\bibitem[\protect\citeauthoryear{Istomin and Pariev}{1994}]{istominStabilityRelativisticRotating1994}
\begin{barticle}
\bauthor{\bsnm{Istomin}, \binits{{\relax Ya}.N.}},
\bauthor{\bsnm{Pariev}, \binits{V.I.}}:
\batitle{Stability of a relativistic rotating electron--positron jet}.
\bjtitle{Monthly Notices of the Royal Astronomical Society}
\bvolume{267}(\bissue{3}),
\bfpage{629}--\blpage{636}
(\byear{1994})
\doiurl{10.1093/mnras/267.3.629}
\end{barticle}
\endbibitem

\bibitem[\protect\citeauthoryear{Istomin and Pariev}{1996}]{istominStabilityRelativisticRotating1996}
\begin{barticle}
\bauthor{\bsnm{Istomin}, \binits{{\relax Ya}.N.}},
\bauthor{\bsnm{Pariev}, \binits{V.I.}}:
\batitle{Stability of a relativistic rotating electron-positron jet: Non-axisymmetric perturbations}.
\bjtitle{Monthly Notices of the Royal Astronomical Society}
\bvolume{281}(\bissue{1}),
\bfpage{1}--\blpage{26}
(\byear{1996})
\doiurl{10.1093/mnras/281.1.1}
\end{barticle}
\endbibitem

\bibitem[\protect\citeauthoryear{{Wei-Ke} et~al.}{2005}]{wei-keDependenceAverageLorentz2005}
\begin{barticle}
\bauthor{\bsnm{{Wei-Ke}}, \binits{A.}},
\bauthor{\bsnm{{Xi-Jun}}, \binits{Q.}},
\bauthor{\bsnm{{Chun-Hua}}, \binits{S.}},
\bauthor{\bsnm{{Zhi-Yuan}}, \binits{Z.}}:
\batitle{Dependence of the {{Average Lorentz Factor}} on {{Temperature}} in {{Relativistic Plasmas}}}.
\bjtitle{Chinese Physics Letters}
\bvolume{22}(\bissue{5}),
\bfpage{1176}
(\byear{2005})
\doiurl{10.1088/0256-307X/22/5/042}
\end{barticle}
\endbibitem

\bibitem[\protect\citeauthoryear{Stenson et~al.}{2017}]{stensonDebyeLengthPlasma2017}
\begin{barticle}
\bauthor{\bsnm{Stenson}, \binits{E.V.}},
\bauthor{\bsnm{{Horn-Stanja}}, \binits{J.}},
\bauthor{\bsnm{Stoneking}, \binits{M.R.}},
\bauthor{\bsnm{Pedersen}, \binits{T.S.}}:
\batitle{Debye length and plasma skin depth: Two length scales of interest in the creation and diagnosis of laboratory pair plasmas}.
\bjtitle{Journal of Plasma Physics}
\bvolume{83}(\bissue{1}),
\bfpage{595830106}
(\byear{2017})
\doiurl{10.1017/S0022377817000022}
\end{barticle}
\endbibitem

\bibitem[\protect\citeauthoryear{Gordienko and Pukhov}{2005}]{gordienko2005scalings}
\begin{botherref}
\oauthor{\bsnm{Gordienko}, \binits{S.}},
\oauthor{\bsnm{Pukhov}, \binits{A.}}:
Scalings for ultrarelativistic laser plasmas and quasimonoenergetic electrons.
Physics of Plasmas
\textbf{12}(4)
(2005)
\doiurl{10.1063/1.1884126}
\end{botherref}
\endbibitem

\bibitem[\protect\citeauthoryear{}{}]{Juwels}
\begin{botherref}
Gauss centre: \href{http://www.gauss-centre.eu}{http://www.gauss-centre.eu}
\end{botherref}
\endbibitem

\bibitem[\protect\citeauthoryear{Pukhov}{2016}]{pukhovParticleInCellCodesPlasmabased2016}
\begin{botherref}
\oauthor{\bsnm{Pukhov}, \binits{A.}}:
Particle-{{In-Cell Codes}} for {{Plasma-based Particle Acceleration}}.
CERN Yellow Reports,
181
(2016)
\doiurl{10.5170/CERN-2016-001.181}
\end{botherref}
\endbibitem

\bibitem[\protect\citeauthoryear{Volokitin et~al.}{2023}]{volokitinOptimizedEventGenerator2023}
\begin{barticle}
\bauthor{\bsnm{Volokitin}, \binits{V.}},
\bauthor{\bsnm{Magnusson}, \binits{J.}},
\bauthor{\bsnm{Bashinov}, \binits{A.}},
\bauthor{\bsnm{Efimenko}, \binits{E.}},
\bauthor{\bsnm{Muraviev}, \binits{A.}},
\bauthor{\bsnm{Meyerov}, \binits{I.}}:
\batitle{Optimized event generator for strong-field {{QED}} simulations within the hi-{$\chi$} framework}.
\bjtitle{Journal of Computational Science}
\bvolume{74},
\bfpage{102170}
(\byear{2023})
\doiurl{10.1016/j.jocs.2023.102170}
\end{barticle}
\endbibitem

\bibitem[\protect\citeauthoryear{Song et~al.}{2022}]{songDensePolarizedPositrons2022}
\begin{barticle}
\bauthor{\bsnm{Song}, \binits{H.-H.}},
\bauthor{\bsnm{Wang}, \binits{W.-M.}},
\bauthor{\bsnm{Li}, \binits{Y.-T.}}:
\batitle{Dense {{Polarized Positrons}} from {{Laser-Irradiated Foil Targets}} in the {{QED Regime}}}.
\bjtitle{Physical Review Letters}
\bvolume{129}(\bissue{3}),
\bfpage{035001}
(\byear{2022})
\doiurl{10.1103/PhysRevLett.129.035001}
\end{barticle}
\endbibitem

\bibitem[\protect\citeauthoryear{Zhuang et~al.}{2023}]{zhuangLaserdrivenLeptonPolarization2023}
\begin{barticle}
\bauthor{\bsnm{Zhuang}, \binits{K.-H.}},
\bauthor{\bsnm{Chen}, \binits{Y.-Y.}},
\bauthor{\bsnm{Li}, \binits{Y.-F.}},
\bauthor{\bsnm{Hatsagortsyan}, \binits{K.Z.}},
\bauthor{\bsnm{Keitel}, \binits{C.H.}}:
\batitle{Laser-driven lepton polarization in the quantum radiation-dominated reflection regime}.
\bjtitle{Physical Review D}
\bvolume{108}(\bissue{3}),
\bfpage{033001}
(\byear{2023})
\doiurl{10.1103/PhysRevD.108.033001}
\end{barticle}
\endbibitem

\end{thebibliography}

\end{document}